\documentclass[runningheads]{llncs}
\usepackage{cite}
\usepackage{amsmath,amssymb,amsfonts}
\usepackage{algorithmic}
\usepackage{graphicx}
\usepackage{textcomp}
\usepackage{xcolor}
\usepackage{comment}
\def\BibTeX{{\rm B\kern-.05em{\sc i\kern-.025em b}\kern-.08em
    T\kern-.1667em\lower.7ex\hbox{E}\kern-.125emX}}
\begin{document}

\title{Developing and Operating Artificial Intelligence Models in Trustworthy Autonomous Systems
}

\author{Silverio Martínez-Fernández \inst{1}\orcidID{0000-0001-9928-133X}
\and
Xavier Franch\inst{1}\orcidID{0000-0001-9733-8830} 
\and
Andreas Jedlitschka\inst{2}
\orcidID{0000-0003-3590-6331}
\and
Marc Oriol\inst{1}
\orcidID{0000-0003-1928-7024}
\and
Adam Trendowicz\inst{2}
%\orcidID{2222--3333-4444-5555}
}

\institute{Universitat Politècnica de Catalunya - BarcelonaTech\\
%, Jordi Girona 1-3, 08030, Barcelona, Spain\\
\email{\{silverio.martinez, xavier.franch, marc.oriol\}@upc.edu} \and
Fraunhofer IESE\\
\email{\{andreas.jedlitschka, adam.trendowicz\}@iese.fraunhofer.de}}

\authorrunning{S.  Martínez-Fernández et al.}

\maketitle

\begin{abstract}
Companies dealing with Artificial Intelligence (AI) models in Autonomous Systems (AS) face several problems, such as users’ lack of trust in adverse or unknown conditions, gaps between software engineering and AI model development, and operation in a continuously changing operational environment. This work-in-progress paper aims to close the gap between the development and operation of trustworthy AI-based AS by defining an approach that coordinates both activities. We synthesize the main challenges of AI-based AS in industrial settings. We reflect on the research efforts required to overcome these challenges and propose a novel, holistic DevOps approach to put it into practice. We elaborate on four research directions: (a) increased users' trust by monitoring operational AI-based AS and identifying self-adaptation needs in critical situations; (b) integrated agile process for the development and evolution of AI models and AS; (c) continuous deployment of different context-specific instances of AI models in a distributed setting of AS; and (d) holistic DevOps-based lifecycle for AI-based AS. %An approach supporting the continuous delivery of evolving AI models and their operation in AI-based AS under adverse conditions would support companies in increasing users’ trust and providing guidance regarding learning-enabled components in AS.
\end{abstract}

\section{Introduction}
Nowadays, Autonomous Systems (AS) are prevalent in many domains: from smart mobility (autonomous driving) and Industry 4.0 (autonomous factory robots) to smart health (autonomous diagnosis systems). %In fact, Gartner identifies AS as one of the top ten strategic technological trends of 2020\cite{gartner2020top}.
One crucial enabler of this success is the emergence of sophisticated Artificial Intelligence (AI) techniques boosting the ability of AS to operate and self-adapt in increasingly complex and dynamic environments. 

Recently, there have been multiple research efforts to understand diverse AS quality attributes (e.g., trustworthiness, safety) with learning-enabled components and how to certify them%\cite{ma2018secure}
. Examples range from the use of declarations of conformity from AI service providers for examination by their consumers \cite{arnold2019factsheets}, to the creation of new standards like SOTIF to consider AS in which learning-enabled components make decisions \cite{gharib2018safety}. This is due to the assumption that the behaviour of such learning-enabled components cannot be guaranteed by neither current software development processes nor by validation and verification approaches \cite{Borg2019}, but they are a starting point \cite{ozkaya2020really}. Certainly, despite being software systems, AI-based AS have different properties than traditional software systems \cite{ozkaya2020really}. In this way, recent research has focused on software engineering best practices for AI-based systems %\cite{khomh2018software, serban2020adoption}
\cite{serban2020adoption}, and specifically for AS \cite{aniculaesei2018toward}. 

In this paper, we elaborate on a novel approach pushing forward the idea that AI-based AS will benefit from an integrated development and operation approach driven by the concept of trustworthiness.

\section{Challenges of developing and operating AI-based AS}

The potential benefits that AI-based AS can provide to their stakeholders, such as data-driven evolution and autonomous behavior \cite{eu20}, have their counterpart in several major challenges that act as impediments to their adoption.

\textit{Increase users’ trust in AI-based AS}. AS could become prevalent in many aspects of people’s life. Correspondingly, trustworthiness on AI is a prerequisite for the uptake of AI-based AS, so that users allow its integration into their life.
Indeed, the European Commission is working on a regulatory framework leading to a unique “ecosystem of trust” \cite{eu20}. Users' trust in AI-based AS is threatened by two reasons. On the one hand, key qualities of AI models (e.g., functional safety, security, reliability, or fairness of decisions) must be guaranteed in all possible (often unanticipated) scenarios. Trustworthiness is a highly complex concept and its achievement poses a number of challenges both during development and operation of AI-based AS. In AI development, it is key to address data quality since the very beginning, in order to build trust into the AI system. At operation, a trustworthy AI system must behave as expected, particularly in unanticipated situations and in case of malfunctioning of any kind, it does not behave in any unwanted, especially harmful, ways. On the other hand, users' trust requires informing users about how the decisions are derived by the AI models. For instance, understanding the reasons behind predictions is a determinant in achieving trust, which is fundamental if one plans to take action based on a prediction, or when choosing whether to deploy a new model \cite{ribeiro2016should}. Despite the widespread adoption of AI technologies, organizations find AI models to be black-boxes and non-transparent \cite{akkiraju2020characterizing,ribeiro2016should}. %Decision understanding is especially critical and justifies the emergence of an important area of research, namely explainable AI (XAI). %\cite{samek2019explainable}.

\textit{Align Software Engineering (SE) and AI model development processes}. Traditional software development lifecycle methodologies fall short when managing AI models, as AI lifecycle management has a number of differences from traditional software development lifecycle \cite{ozkaya2020really} (e.g., %agile vs. waterfall;
requirements-driven vs. data-driven) \cite{akkiraju2020characterizing}. Furthermore, as reported by Atlassian and Microsoft \cite{kim2017data}, it is important to explore the role of humans from different areas (e.g., software and AI model development) in a complex modern industrial environment where AI-based systems are developed. Therefore, an integration of software and AI processes becomes necessary for AI-based AS \cite{heck2019seml, santhanam2019engineering}. In addition, this integration shall support continuously adapting AI models based on evolving users’ needs and changing environments (i.e., to cope with concept drift) \cite{saltz2017comparing}.

\textit{Context-dependent AS and AI model version control and deployment}. Learning-enabled components of AI-based AS are trained and tested with various combinations of parameters, using different data sets, customized to personalized environments, and even solved by diverse algorithms and solutions. AI model deployment challenges in industrial settings include training-serving skew, difficulties in designing the serving infrastructure, and difficulties with training at the edge \cite{lwakatare2020large}. These challenges have motivated the concepts of AIOps and MLOps in the grey literature, and best practices like versioning \cite{serban2020adoption}. However, solutions or frameworks supporting the management of multiple instances of AI models deployed in different context-aware environments are scarce. Context is the missing piece in the AI lifecycle \cite{garcia2018context}.

\textit{Closing the gap between the development and operation phases of the AI-based AS}.  %AI-based AS are highly dependent on the underlying data. So, consistency, accuracy, and completeness of data is essential. There are multiple data management challenges that must be considered, from its collection to its post-deployment (e.g., feedback loops) \cite{raj2019data}.
Nguyen-Duc et. al. \cite{nguyen2020multiple} summarize the engineering challenges for developing and operating AI systems into seven categories: requirements, data management, model design and implementation, model configuration, model testing, evaluation and deployment, and processes and practice. Kästner et. al. argue on the need of improving the educational skills covering the whole software and AI lifecycle \cite{kastner2020teaching}. Proposals to adopt DevOps have also been presented for AS, such as in %spacecraft flight software \cite{heistand2019devops} or
autonomous vehicles \cite{banijamali2019kuksa}. These AI-based AS have some characteristics that differentiate them from other types of software systems, e.g., AI models are usually deployed as part of embedded systems, which require additional techniques for continuous deployment, such as Over-the-Air updates \cite{banijamali2019kuksa}. However, as aforementioned, none of the these approaches have a focus on trustworthiness nor context-specific AI model deployment. Hence, we consider that the development and operation of trustworthy AI-based AS require a holistic approach including enabling iterative cycles for reliably training, adapting, maintaining, and operating the AI model in AS.

\section{Research directions}

In this section, we discuss four research directions to address the previous challenges about development and operation approaches for AI-based AS.

\textbf{Direction 1: Increasing users' trust in AI-based AS by means of transparent real-time monitoring of AI models trustworthiness}. An AS requires self-adaptation to rapidly and effectively protect their critical capabilities from disruption caused by adverse events and conditions. Hence, this direction focuses on continuous self-monitoring in operation based on a set of indicators aggregated into a \emph{trustworthiness score} (TWS). The TWS is a high-level indicator summarizing the level of trustworthiness of an AI model. It is a mean to consider evolving users’ needs and changing operational environments and can be used to guide the self-adaptation of AI models in operations, following the commonly used MAPE-K loop \cite{computing2003architectural}. The high-level TWS is broken down into specific and measurable aspects of AI models at operation (not fully covered in software quality measurement standards \cite{siebert2020towards}): security, dependability, integrity, and reliability. This vision on the computation of indicators of AI trustworthiness is shared by C. Green \cite{ericsson2020} and emerging standard groups \cite{iso2020ai}. In our case, TWS aspects are measured from contextual data collected by AS (e.g., via sensors) and monitoring the AI models (e.g., accuracy). The TWS triggers an action when outliers are predicted or identified. Then, the AS self-adapts.
%For example, Kumar et. al. consider the sum of three components: ethics of algorithms, ethics of data, and ethics of practice \cite{kumar2020trustworthy}. Another component is explainability, as it has been explored in medicine \cite{holzinger2017we}.

These types of ‘scores’ are attractive for practitioners to transparently communicate and understand the real-time the status of AI models and alert the development team (or even the end-user) in case of potential threats/risks. Indeed, some companies have proposed scores for other qualities, such as Google is quantifying testing issues to decrease machine learning technical debt \cite{breck2017ml}. Furthermore, a Gartner report considers tracking AI-based systems to enable transparency as a critical lesson learned from early AI projects: “If [an AI practitioner] predicts that something will fail, the immediate question someone asks is ‘Why is it going to fail?’  [...] having the transparency will help him explain how he came to that conclusion” \cite{gartner2018lessons}.

\textbf{Direction 2: Integrating the development and evolution of AI models and AS}.
Reinforcing the adaptation capabilities of AI models during their development requires highly iterative engineering processes, including the data science and AI model development processes of building, evaluating, deploying, and maintaining AI models, and software systems based upon them. An AI-based AS has two main assets: the AS, and the AI model(s) embedded in a learning-enabled component. Often, due to the methodological gap between the AI model development process and software engineering, these two assets are developed in two parallel, but independent cycles. In other words, there is a need for a seamless integration of these processes.

In addition, for two decades, agile principles have been successfully applied for the rapid and flexible development of high-quality software products \cite{gustavsson2016benefits}, whereas AI model development has been guided by relatively abstract and inflexible processes \cite{saltz2017comparing}. It is time to break these silos. Therefore, this direction proposes an integrated process with coordinated communication between the AI model development and SE teams to develop and evolve AI models for AS. This integrated process both adopts the principles of agile software development and integrates them with existing and well-known AI model development processes, such as the Cross-industry Standard Process for Data Mining \cite{shearer2000crisp, studer2020towards}. This is a complex task requiring the integration of different activities, roles, and multiple existing methodologies. Initial efforts in this direction are the Team Data Science Process from Microsoft \cite{microsoft2020tdsp} and tool support like IBM Watson Studio \cite{ibm2020watson}.

\textbf{Direction 3: Providing intelligent and context-aware techniques to deploy updated AI models in AS instances}. With key data from operation, AI models are continuously evolving to address trust-related threats/risks. Even when AS have the same characteristics (e.g., two autonomous vehicles of the same type), their stakeholders (e.g., owners/passengers with diverse driving style, mood) and environments (e.g., weather, traffic) vary. This is why learning-enabled components of AS need to continuously adapt or even evolve to their users and contexts. Thus, the number of instances may evolve from only one instance (e.g., in an Industry 4.0 machine infrastructure) to thousands of instances (e.g., autonomous vehicles in a smart city). A critical challenge is that often there are no clearly distinguishable components or parts of an AI model that can be analyzed regarding commonalities and variabilities.

This direction aims to facilitate a context-specific deployment of AI models in diverse AS. From the SE perspective, available research on managing software product lines might be reused to manage variabilities and commonalities among contexts in which AS are operating in and consequently among the data they collect evolve on \cite{capilla2019software}. From the data science and AI perspective, the research on transfer learning \cite{torrey2010transfer} and active learning \cite{settles2012active} provide valuable strategies for sharing knowledge between various contexts and incrementally evolving AI models. 

\textbf{Direction 4: Bringing together the development and operation of AI models in trustworthy AS into a holistic lifecycle with tool support}. The SE community is currently researching the application of SE for AI-based systems \cite{serban2020adoption}. However, challenges regarding maintenance and deployment of AI models still remain \cite{kastner2020teaching, lwakatare2020large}. For instance, a survey conducted by SAS revealed that less than 50\% of AI models get deployed and for those that do, it takes more than three months to complete the deployment \cite{sas2019get}. In this context, it becomes necessary "to accelerate getting AI/ML models into production” \cite{broda2013enterprise}.

To keep the development and operation of AI models interconnected, this direction proposes an effective DevOps holistic lifecycle for the production of trustworthy AI-based AS. Since companies typically cannot afford drastic changes of their methodology, a plug-and-play process components are of high importance. This direction includes the development of AI-specific, independent, and loosely coupled software components (ready to be integrated into companies’ development and operational environments) for the three directions we presented above.

\section{A DevOps Approach to Develop and Operate AI Models in AS}

To address the four research directions, we propose an integrated approach bringing the concept of DevOps to AI models of AS. DevOps “integrates the two worlds of development and operation, using automated development, deployment, and infrastructure monitoring” \cite{ebert2016devops}. It is an organizational shift in which cross-functional teams work on continuous operational feature deliveries \cite{ebert2016devops}. 

Additionally, in development, the approach also aims to close the gap between data science/AI model development and SE, as shown in Fig. \ref{fig:devops}. Currently, frameworks are mainly based on the AI models lifecycle in isolation, and have not addressed the issue of integrating the AI pipeline in the DevOps cycle. The proposed approach addresses the inherent challenges of managing multiple instances of models deployed in a distributed set of autonomous systems, such as deploying multiple models in a distributed and heterogeneous environment; and collecting data from multiple sources to assess the behaviour of those models. As depicted in Fig. \ref{fig:devops}, we propose two cycles running in parallel:

\begin{figure}[!tb]
\centering
\includegraphics[width=0.75\textwidth]{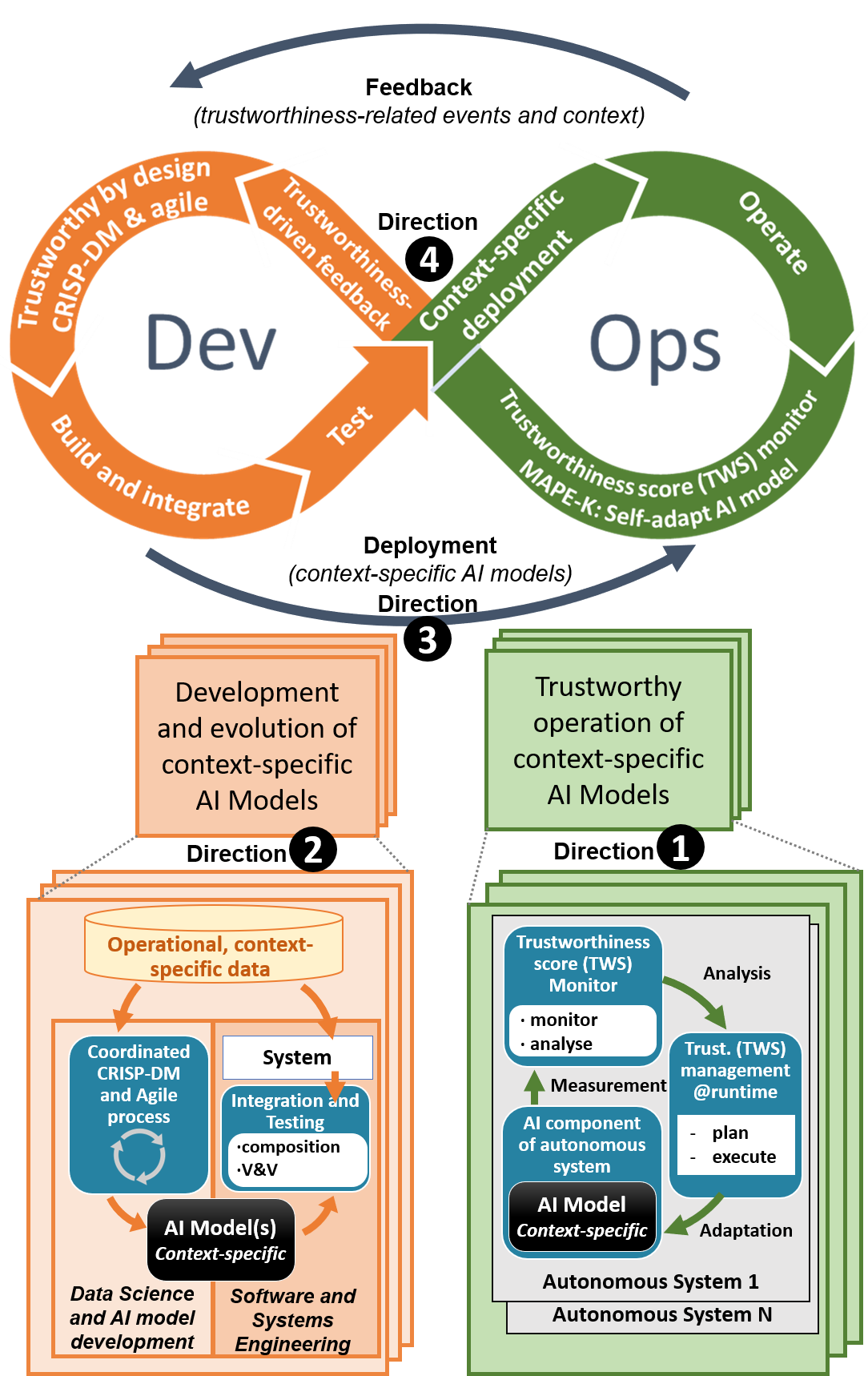}
\caption{Integrated approach for the development and operation of AI-based AS.}
\label{fig:devops}
\end{figure}

\textit{1. An operation cycle (Ops) in the form of a MAPE-K loop\cite{computing2003architectural} ensuring trustworthy behaviour of context-specific AI models by means of continuous self-adaptation capabilities and resilience approaches }(Direction 1 in Fig. \ref{fig:devops}). It enables AI-based AS to rapidly and effectively protect their critical capabilities from disruption caused by adverse events and conditions (e.g., enable an actuator to avoid a collision in an autonomous vehicle). Therefore, it is crucial to continuously monitor the behaviour of the AS (in particular, with respect to trustworthiness, including security, dependability, integrity, and reliability threats caused by the AI model) in new situations such as e.g., energy supply problems, malfunctioning of some component, plan for new or improved functionalities, and finally adapt their individual operation behaviour during execution.

\textit{2. An integrated AI Development cycle (Dev) that adopts agile practices for the iterative development and evolution of context-specific AI models} (Direction 2 in Fig. \ref{fig:devops}). It includes continuous improvement of the AI models based on operational data and enriched by experts’ knowledge. The development, maintenance and evolution of AI-based AS relies on the collaboration of AI model development and software engineering teams. The engineers involved in the maintenance of AI models in the AS are supported by (1) the TWS from operation that may require updating AI models, (2) key data collected regarding the trustworthiness of the AS and its environment, and (3) tools to prepare and enrich the collected data. The software and system engineering team integrate the AI models into the  AS and implement the infrastructure to monitor the trust-related risks of the AS during its operation to ensure the trustworthiness of the entire AS.

These two cycles are integrated into a holistic DevOps approach for trustworthy AI-based AS (Direction 4 in Fig. \ref{fig:devops}), which acts as a bridge between the \textit{Dev} and \textit{Ops} cycles as follows. First, \textit{Dev} deploys the AS with its incorporated context-specific AI models (Direction 3 in Fig. \ref{fig:devops}). AI models are deployed in a distributed context-specific AS. Such deployment requires specific activities, such as context identification of the AS to deploy the correct context-specific AI model instance, Over-The-Air updates to enact the deployments, and validation of AI models within the running context against predefined capabilities and bounds for data gathering or executing axioms. Second, during \textit{Ops}, a continuous cycle gathers trustworthiness-related events and context-specific feedback data from AI models at operation and sends them to development (feedback arrow of Fig. \ref{fig:devops}). Key quality objectives for trustworthy AS are identified, e.g., effectiveness and efficiency of the AI model or environmental and social effects. The monitoring infrastructure provides feedback to manage potential risks in operation (self-adaptation of Direction 1) and in development (evolution of the AI model of Direction 2). Then, the approach starts over again, and \textit{Dev} uses the feedback to evolve the AI models and deploy them back in existing instances and eventually in the system functionalities. 

\section{Concluding remarks}

We identified the challenges that need to be overcome to develop and operate trustworthy AI-based AS. We discussed four research directions to address such challenges. Finally, we present a holistic DevOps approach bringing together the development and operation of AI models in AS, aiming to increase users’ trust.%, which are becoming pervasive in our society.

Several factors may hinder the successful application of these research directions. First, the need to ensure the reliability of the TWS for different purposes and scenarios, e.g., need for customizable TWS, as well as recognizing that some aspects are challenging to be measured (e.g., ethics). Second, many companies already have pipelines to create AI-based AS based on a unique perspective (either data science or system engineering) and may be unwilling to adopt a unified approach. Third, even though being minimal, the iterative feedback cycles trigger discussions about ethics because users might be monitored (even though complying with their agreement for data collection and current legislation), and processing huge amounts of data leaves a “carbon footprint”.

Future work consists of exploring how the elements of this integrated approach can be exploited or customised to support existing AI-based AS.

\section*{Acknowledgment}
This work has been partially supported by the Beatriz Galindo programme (BGP18/00075) and the Catalan Research Agency (AGAUR, contract 2017 SGR 1694).

%\newpage

\bibliographystyle{IEEEtran}
\bibliography{IEEEabrv,referencesfile}

\vspace{12pt}

\end{document}